\documentclass[reprint,showpacs,preprintnumbers,amsmath,amssymb,aps,pre]{revtex4-1}

\usepackage{graphicx}
\usepackage{dcolumn}
\usepackage{bm}
\usepackage[OT4]{fontenc}

\begin{document}

\title{Interplay between charge and magnetic orderings in the zero-bandwidth\\
limit of the extended Hubbard model for strong on-site repulsion}
\author{Konrad Kapcia}%
    \email{corresponding author; e-mail: \url{kakonrad@amu.edu.pl}}
\affiliation{%
Faculty of Physics, Adam Mickiewicz University, Umultowska 85, 61-614 Pozna\'n, Poland
}%
\author{Waldemar K\l{}obus}
\affiliation{%
Faculty of Physics, Adam Mickiewicz University, Umultowska 85, 61-614 Pozna\'n, Poland
}
\author{Stanis\l{}aw Robaszkiewicz}%
\affiliation{%
Faculty of Physics, Adam Mickiewicz University, Umultowska 85, 61-614 Pozna\'n, Poland
}%

\date{May 1, 2011}

\begin{abstract}
A simple effective model of charge ordered and (or) magnetically ordered insulators is studied. The tight binding Hamiltonian analyzed consists of (i)~the effective on-site interaction $U$, (ii)~the intersite density-density interaction $W$ and (iii)~intersite magnetic exchange interaction $J^z$ (or $J^{xy}$) between nearest-neighbors.
The intersite interaction are treated within the mean-field approximation.
One shows that the systems considered can exhibit very interesting multicritical behaviors, including among others bicritical, tricritical, tetracritical and critical end points.  The analysis of the model has been performed for an arbitrary electron concentration as well as an arbitrary chemical potential in the limit of strong on-site repulsion (\mbox{$U\rightarrow+\infty$}). The phase diagrams obtained in such a~case are shown to consist of at least 9 different states, including four homogenous phases: nonordered (NO), ferromagnetic (F), charge ordered (CO), ferrimagnetic (intermediate, I) and five types of phase separation: \mbox{NO -- NO}, \mbox{F -- NO}, \mbox{F -- F}, \mbox{CO -- F}, \mbox{CO -- I}.
\end{abstract}

\pacs{71.10.Fd, 71.45.Lr, 75.30.Fv, 64.75.Gh, 71.10.Hf}
\maketitle


\section{Introduction}

Electron charge orderings and their interplay with magnetism are relevant to broad range of important materials such as manganites, multiferroics and other strongly correlated electron systems \cite{MRR1990,GL2003,BK2008}. In this report we discuss an effective model of charge ordered and (or) magnetically ordered insulators.

We consider  the following model Hamiltonian of a~fermion lattice system in the atomic limit
\begin{eqnarray}
\label{row:hamiltonian}
\hat{H} & = & \sum_i{U\hat{n}_{i\uparrow}\hat{n}_{i\downarrow}} - \sum_i{\mu\hat{n}_{i}} + \\
& + & \frac{W}{2}\sum_{\langle i,j\rangle}{\hat{n}_{i}\hat{n}_{j}}
- 2J\sum_{\langle i,j\rangle}{\hat{s}^z_{i}\hat{s}^z_{j}}, \nonumber
\end{eqnarray}
where $\hat{c}^{+}_{i\sigma}$ denotes the creation operator of an electron with spin $\sigma$ at the site $i$,
\mbox{$\hat{n}_{i\sigma}=\hat{c}^{+}_{i\sigma}\hat{c}_{i\sigma}$},
\mbox{$\hat{n}_{i}=\sum_{\sigma}{\hat{n}_{i\sigma}}$}, \mbox{$\hat{s}_i^z=(1/2)(\hat{n}_{i\uparrow}-\hat{n}_{i\downarrow})$}.
$U$, $W$ and $J$ are the interaction parameters (on-site and intersite between the nearest neighbors). $\mu$ is the chemical potential, depending on the concentration of electrons
\begin{equation}
\mbox{$n = \frac{1}{N}\sum_{i}{\left\langle \hat{n}_{i} \right\rangle}$},
\end{equation}
with $n\in[0,2]$, $N$ is the total number of lattice sites. Our denotations:
\mbox{$n_Q=\frac{1}{2}(n_A-n_B)$}, \mbox{$n_\alpha=\frac{2}{N}\sum_{i \in \alpha}\langle \hat{n}_i \rangle$},
\mbox{$m=\frac{1}{N}\sum_i\langle \hat{s}^z_i\rangle$}, \mbox{$m_Q=\frac{1}{2}(m_A-m_B)$}, \mbox{$m_\alpha=\frac{2}{N}\sum_{i\in\alpha}\langle \hat{s}^z_i\rangle$}, and \mbox{$\alpha=A,B$} labels the sublattices.
$W_0 = zW$ and $J_0 = zJ$, where $z$ is the number of the nearest neighbors. Only the two-sublattice orderings on the alternate lattices are considered in this report.

The interactions $U$, $W$, and $J$ will be treated as effective ones and  be assumed  to include all the possible contributions and renormalizations.
One should notice that in absence of external magnetic field the ferromagnetic (\mbox{$J>0$}) interactions are simply mapped onto the antiferromagnetic cases (\mbox{$J<0$}) by redefining the spin direction on one sublattice in lattices decomposed into two interpenetrating sublattices. Thus, we restrict ourselves to the case \mbox{$J>0$}.

In the analysis we have adopted a~variational approach (VA), which treats the on-site interaction term ($U$) exactly and the intersite interactions ($W$, $J$) within the mean-field approximation (MFA).

For the model (\ref{row:hamiltonian})  a ground state phase diagram as a function of $\mu$ \cite{BS1986} (exact results) and special cases at \mbox{$T\geq0$} (in VA) such as \mbox{$W=0$} \cite{KKR2010a}, \mbox{$J=0$} \cite{MRC1984,KKR2010b,KR2011} and \mbox{$n=1$} \cite{R1979} have been investigated till now.

\begin{table}
    \caption{\label{tab:table1}Definitions of homogeneous phases and PS states.}
    \begin{ruledtabular}
\begin{tabular}{|p{0.09\linewidth} p{0.43\linewidth}||p{0.1\linewidth}  p{0.07\linewidth} p{0.07\linewidth}|}
Phase &  Order parameters &  State  & $n_{\pm}$ & $n_{\mp}$ \\
\hline
         &                       & PS2  & CO & NO  \\

 CO&    $n_Q\neq0$, $m=0$, $m_Q=0$       & PS3 & NO & NO    \\

 F&     $n_Q=0$, $m\neq0$, $m_Q=0$       & PS4 & F & NO   \\

 I&     $n_Q\neq0$, $m\neq0$, $m_Q\neq0$    & PS5 & I & CO   \\

 NO&    $n_Q=0$, $m=0$, $m_Q=0$       & PS6 & F & CO    \\

         &                       & PS7 & F & F    \\
\end{tabular}
    \end{ruledtabular}
\end{table}

Within the VA the intersite interactions are decoupled within the MFA, which allows us to find a~free energy per site $f(n)$.
One can also calculate the averages: $n$, $n_Q$, $m$ and $m_Q$, what gives a set of four self-consistent equations (for homogeneous phases).
This set for $T\geq0$ is solved numerically and one obtains $n_Q$, $m$, $m_Q$
and $n$ (or $\mu$) when $\mu$ (or $n$) is fixed. It is important to find a~solution corresponding to the lowest energy.

The phase separation (PS) is a state in which  two domains with different electron concentration exist
(coexistence of two homogeneous phases). The energy of PS states is given by
\begin{equation}
f_{PS}(n_{+},n_{-}) = k f_{+}(n_{+}) + (1-k) f_{-}(n_{-}),
\end{equation}
where $E_{\pm}(n_{\pm})$ are values of a free energy at $n_{\pm}$ corresponding to the
lowest  energy homogeneous solutions and
\mbox{$k  = \frac{n - n_-}{n_+ - n_-}$}
is a fraction of the system with density $n_+$.

In the report we have used the following convention. A~second (first) order transition is a~transition between homogeneous phases  with a~(dis-)continuous change of the order parameter at the transition temperature. A~transition between homogeneous phase and PS state is symbolically named as a~``third order'' transition. During this transition a~size of one domain in the PS state decreases continuously to zero at the~transition temperature.  We have also distinguished a~second order transition between two PS states, at which a~\mbox{continuous} change of the order parameter in both domains takes place.

\begin{figure}
    \centering
        \includegraphics[width=0.43\textwidth]{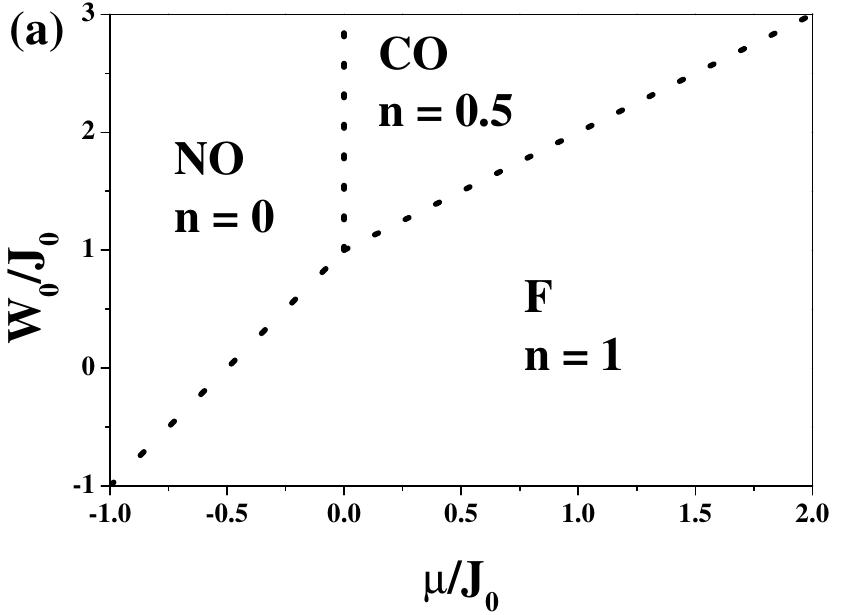}\\
        \includegraphics[width=0.43\textwidth]{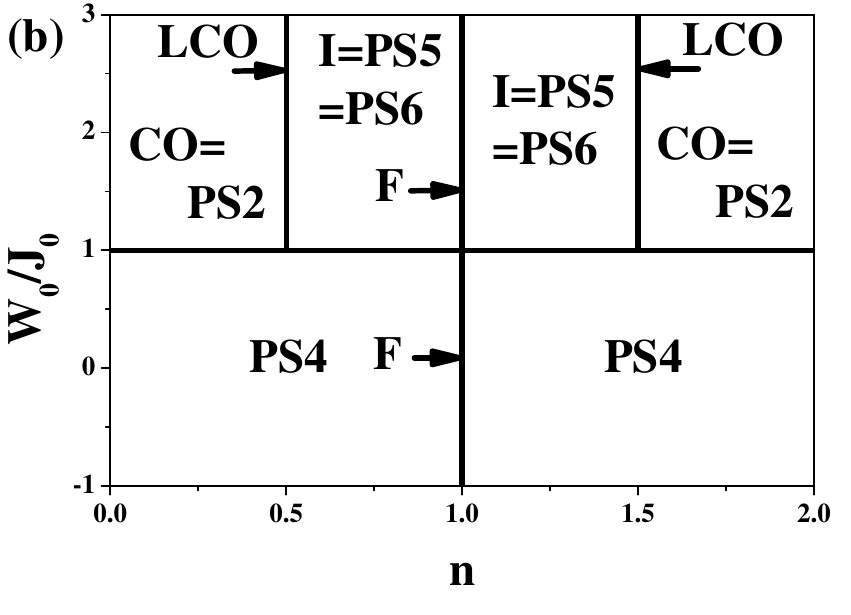}
        \caption{Ground state phase diagrams in the \mbox{$U\rightarrow+\infty$} limit: (a)~as a function of $\bar{\mu}/J_0$ and (b)~as a~function of $n$. Dotted lines denote discontinuous transitions (on the panel (a)).}
        \label{rys:GS}
\end{figure}

In this report we present the VA results for the model (\ref{row:hamiltonian}) in the limit \mbox{$U\rightarrow+\infty$}. This case corresponds to the subspace excluding the double occupancy
of sites (by electrons for \mbox{$n < 1$} or holes for \mbox{$n > 1$}). The obtained phase diagrams (for fixed $n$) are symmetric with respect to half-filling (\mbox{$n=1$}) because of the particle--hole symmetry of the Hamiltonian (\ref{row:hamiltonian}).
The phases and states, which can exist on the phase diagrams of model considered, are collected in Table~\ref{tab:table1}.

\section{Results and discussion for \mbox{$U\rightarrow +\infty$}}

\begin{figure}
    \centering
    \includegraphics[width=0.5\textwidth]{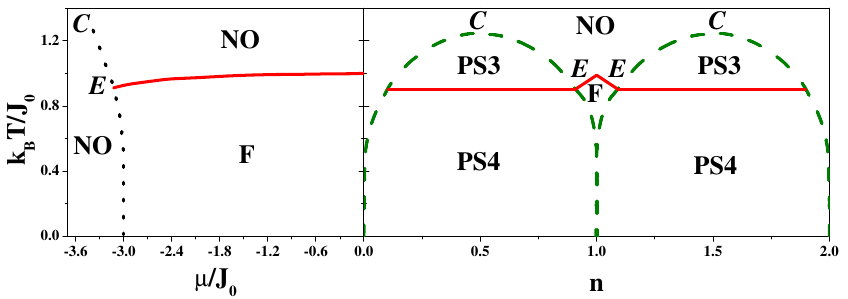}\\
    \caption{Finite temperature phase diagrams for \mbox{$W/J=-5$} (the \mbox{$U\rightarrow+\infty$} limit). Dotted, solid and dashed lines indicate first order, second order and ``third order'' boundaries, respectively. Details in text.}
    \label{rys:fig1}
\end{figure}

\begin{figure}
    \centering
    \includegraphics[width=0.5\textwidth]{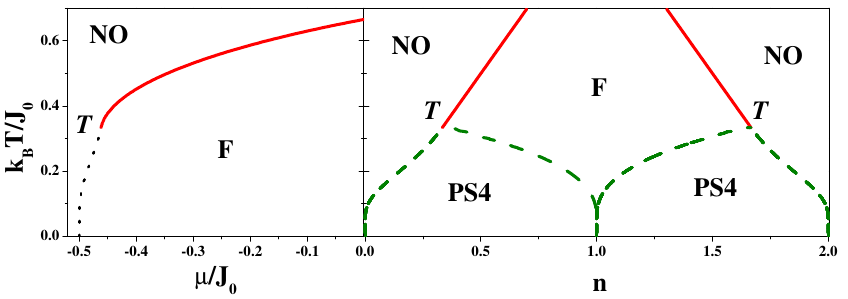}
    \caption{Finite temperature phase diagrams for \mbox{$W/J= 0$} and \mbox{$U\rightarrow+\infty$}. Denotations as in Fig.~\ref{rys:fig1}.}
    \label{rys:fig1a}
\end{figure}

In this section we discuss the behavior of the system in the limit of infinite on-site interaction (\mbox{$U\rightarrow+\infty$}). The ground state diagrams obtained in this limit are shown in Fig.~\ref{rys:GS}. The energies of a few states are degenerated if the system is considered for fixed $n$. However, at finite temperatures the degeneration is removed. LCO is the particular CO phase with \mbox{$n_Q=0.5$} and \mbox{$n=0.5$} or \mbox{$n=1.5$}.

\begin{figure}
    \centering
    \includegraphics[width=0.5\textwidth]{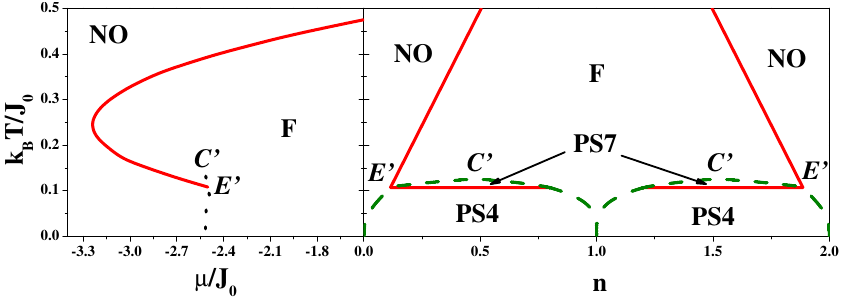}\\
    \caption{Finite temperature phase diagrams for \mbox{$W/J= 0.5$} and \mbox{$U\rightarrow+\infty$}. Denotations as in Fig.~\ref{rys:fig1}.}
    \label{rys:fig2}
\end{figure}

\begin{figure}
    \centering
    \includegraphics[width=0.5\textwidth]{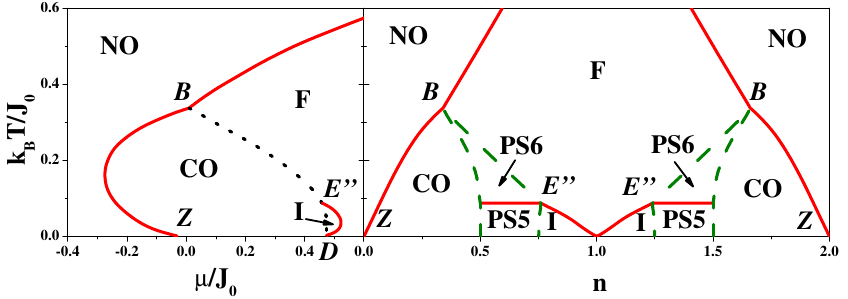}\\
    \caption{Finite temperature phase diagrams for \mbox{$W/J=1.5$} and \mbox{$U\rightarrow+\infty$}. Denotations as in Fig.~\ref{rys:fig1}.}
    \label{rys:fig9}
\end{figure}

The selected finite temperature phase diagrams, obtained for fixed $\mu$ as well as for fixed $n$, are presented in \mbox{Figs.~\ref{rys:fig1}--\ref{rys:fig9}}. One should notice that the first order transitions on the diagrams for fixed $\mu$ corresponds to the ``third order'' transitions and occurrence of PS sates on the diagrams for fixed $n$. The second order transition F-NO occurs at \mbox{$k_BT=J_0|1-n|$} independently of $W/J$.

The labels of the critical points for phase separation are given corresponding to those in \cite{HB1991}, where the Blume-Emery-Griffiths model was considered.
Critical points
indicated on the phase diagrams (descriptions in correspondence to diagrams as a function of $\mu/J_0$, Figs.~\ref{rys:fig1}--\ref{rys:fig9}):
(i)~$T$
-- tricritical point: change of the transition order,
(ii)~$C$, $C'$ -- isolated critical (bicritical end) points: the end of 1st order transition line,
(iii)~$E$, $E'$, $E''$ -- critical end points: connection of two 1st order lines and one 2nd order line,
(iv)~$Z$ -- zero-temperature critical point: the end of 2nd order transition line at \mbox{$T=0$},
(v)~$D$ -- zero-temperature highly degenerate point: the end of both 1st and 2nd order transitions lines at \mbox{$T=0$},
(vi)~$B$ -- bicritical point: connection of one 1st order and two 2nd order lines.

For large negative values of $W/J$, a~second order \mbox{F--NO} line terminates at $E$-point on a~first order line, which itself ends at $C$-point (Fig.~\ref{rys:fig1}). For values of \mbox{$W/J\simeq0$} (Fig.~\ref{rys:fig1a}), the transition lines meet $T$-point, which is connected with a~change of \mbox{F--NO} transition order \cite{KKR2010a}.

With increasing $W/J$ the tricritical
behavior changes into another one with the $C'$-point located inside the F phase (cf. Fig.~\ref{rys:fig2}, for \mbox{$W/J=0.5$}).

For $W/J>1$ (Fig.~\ref{rys:fig9}) two new ordered phases appear on the phase diagrams: charge ordered (CO) phase (\mbox{$n_Q\neq0$}) and intermediate (I) homogeneous phase, where magnetic order and charge order coexist (\mbox{$n_Q\neq0$}, \mbox{$m\neq0$} and \mbox{$m_Q\neq0$}). If \mbox{$W/J=1.5$}
the F and CO phases are separated from the NO phase by two second order lines that connect at $B$-point. The transitions between CO and phases with \mbox{$m\neq0$} (F, I) are first order. The F and I phases are separated by second order line ending at $E''$-point.

\section{Final remarks}

It is seen that a~large variety of phase transition phenomena are present in the simple model incorporating both magnetic and  density degrees of freedom. Derived results are exact in the limit of infinite dimensions, where the MFA treatment of intersite interactions becomes the rigorous one.

Our model in the limit \mbox{$U\rightarrow+\infty$} is equivalent to the Blume-Emery-Griffiths model, which is the general spin \mbox{$S=1$} Ising model with nearest-neighbor interactions and up-down symmetry. Thus obtained diagrams have the similar structure as those presented in Ref.~\onlinecite{HB1991}.

One notice that in the \mbox{$U\rightarrow-\infty$} limit of model (\ref{row:hamiltonian}) the magnetic orderings are completely suppressed and the CO phase (for \mbox{$W>0$}) or the PS: \mbox{NO--NO} (for \mbox{$W<0$}) state can occur. For \mbox{$W=0$} only the NO is stable at any \mbox{$k_BT\geq 0$} in this limit.

Let us point out that in MFA  the following equivalence occurs: \mbox{$2J\sum_{\langle i,j \rangle} {\hat{s}^z_i\hat{s}^z_j} \stackrel{MFA}{\longleftrightarrow} J\sum_{\langle i,j \rangle} {\left(\hat{s}^+_i\hat{s}^-_j+\hat{s}^+_j\hat{s}^-_i\right)}$}.
In both cases the self-consistent equations have the same form and a magnetization along the $z$-axis becomes a magnetization in the $xy$-plane \cite{WK2009}.

\begin{acknowledgments}

K.~K. would like to thank the European Commission and Ministry of Science and Higher Education (Poland) for the partial financial support from European Social Fund -- Operational Programme ``Human Capital'' -- POKL.04.01.01-00-133/09-00 -- ``\textit{Proinnowacyjne kszta\l{}cenie, kompetentna kadra, absolwenci przysz\l{}o\'sci}''.

\end{acknowledgments}


\begin{thebibliography}{10}

\bibitem{MRR1990}
R. Micnas, J. Ranninger, S. Robaszkiewicz, \textit{Rev. Mod. Phys.} \textbf{62}, 113 (1990).
\bibitem{GL2003}
T. Goto, B. L\"uthi, \textit{Adv. Phys.} \textbf{52}, 67 (2003);
E. Dagotto, T. Hotta, A. Moreo, \textit{Phys. Reports} \textbf{344}, 1 (2001).
\bibitem{BK2008}
J. van den Brink, D. I. Khomskii, \textit{J. Phys.: Condens. Matter} \textbf{20}, 434217 (2008).
\bibitem{BS1986}
U. Brandt, J. Stolze, \textit{Z. Phys.}~B \textbf{62}, 433 (1986); J.~J\k{e}drzejewski, \textit{Physica}~A, \textbf{205}, 702 (1994).
\bibitem{KKR2010a}
W. K\l{}obus, K. Kapcia, S. Robaszkiewicz, \textit{Acta. Phys. Pol.}~A \textbf{118}, 353 (2010).
\bibitem{MRC1984}
R. Micnas, S. Robaszkiewicz, K.~A.~Chao, \textit{Phys. Rev.}~B \textbf{29}, 2784 (1984).
\bibitem{KKR2010b}
K.~Kapcia, W.~K\l{}obus, S.~Robaszkiewicz, \textit{Acta. Phys. Pol.}~A \textbf{118}, 350 (2010).
\bibitem{KR2011}
K.~Kapcia, S.~Robaszkiewicz, \textit{J. Phys.: Condens. Matter} \textbf{23}, 105601 (2011); \textbf{23}, 249802 (2011).
\bibitem{R1979}
S. Robaszkiewicz, \textit{Acta Phys. Pol.}~A \textbf{55}, 453 (1979); \textit{Phys. Status Solidi}~(b) \textbf{70}, K51 (1975).
\bibitem{HB1991}
W. Hoston, A. N. Berker, \textit{Phys. Rev. Lett.} \textbf{67}, 1027 (1991); C.~Ekiz, M.~Keskin, \textit{Phys. Rev.}~B \textbf{66}, 054105 (2002).
\bibitem{WK2009}
W. K\l{}obus, Master thesis, Adam Mickiewicz University, Pozna\'n (2009).
\end{thebibliography}
\end{document}